\documentclass[pdflatex,sn-mathphys-num]{sn-jnl}% Math and Physical Sciences Numbered Reference Style
\usepackage{graphicx}%
\usepackage{multirow}%
\usepackage{amsmath,amssymb,amsfonts}%
\usepackage{amsthm}%
\usepackage{mathrsfs}%
\usepackage[title]{appendix}%
\usepackage{xcolor}%
\usepackage{textcomp}%
\usepackage{manyfoot}%
\usepackage{booktabs}%
\usepackage{algorithm}%
\usepackage{algorithmicx}%
\usepackage{algpseudocode}%
\usepackage{listings}%
\usepackage[left]{lineno}

\theoremstyle{thmstyleone}%
\theoremstyle{thmstyletwo}%

\theoremstyle{thmstylethree}%

\begin{document}
% \linenumbers
% \modulolinenumbers[5]

% Emergent homogeneity in online communities without algorithms or preference-based segregation
% \title[Article Title]{Schelling Segregation can Explain the Emergence of Online Echo Chambers}
% \title{Emergent homogeneity in online communities without algorithms or preference-based segregation}
\title{Online Homogeneity Can Emerge Without Filtering Algorithms or Homophily Preferences}
% \title{Online homogeneity can emerge without algorithms or preference-based segregation}


\author*[1]{\fnm{Petter} \sur{Törnberg}}\email{p.tornberg@uva.nl}

\affil*[1]{\orgdiv{ILLC}, \orgname{University of Amsterdam}, \orgaddress{\street{Science Park 900}, \postcode{1098 XH}, \city{Amsterdam}, \country{The Netherlands}}}


\abstract{Ideologically homogeneous online environments --- often described as ``echo chambers'' or ``filter bubbles'' --- are widely seen as drivers of polarization, radicalization, and misinformation. A central debate asks whether such homophily stems primarily from algorithmic curation or users' preference for like-minded peers. This study challenges that view by showing that homogeneity can emerge in the absence of both filtering algorithms and user preferences. Using an agent-based model inspired by Schelling's model of residential segregation, we demonstrate that weak individual preferences, combined with simple group-based interaction structures, can trigger feedback loops that drive communities toward segregation. Once a small imbalance forms, cascades of user exits and regrouping amplify homogeneity across the system. Counterintuitively, algorithmic filtering --- often blamed for ``filter bubbles'' --- can in fact sustain diversity by stabilizing mixed communities. These findings highlight online polarization as an emergent system-level dynamic and underscore the importance of applying a complexity lens to the study of digital public spheres.}




\maketitle

% \begin{quote}
% \small
% \textbf{Teaser:} Homogeneity online can emerge without intent or algorithms, driven by tipping cascades that produce highly sorted systems.
% \end{quote}

\section{Introduction}\label{sec1}
Ideologically homogeneous online environments --- variously termed 'echo chambers' \cite{sunstein2018republic}, 'deliberative enclaves' \cite{mutz2006hearing}, 'cyberbalkanization' \cite{van1996electronic}, or 'filter bubbles' \cite{pariser2011filter} --- have been widely associated with negative social and political outcomes. These include rising political polarization \cite{sunstein2002law}, radicalization \cite{tornberg2024intimate}, the reinforcement of confirmation biases \cite{stroud2010polarization,garrett2009echo}, the proliferation of misinformation \cite{tornberg2018echo}, and the divergence of fundamental worldviews among users \cite{barbera2015tweeting}. While scholars debate the extent and pervasiveness of online homophily \cite{dubois2018echo,bruns2019filter,bakshy2015exposure}, as well as its social implications \cite{guess2023social,bail2018exposure, tornberg2024intimate, nyhan2023like}, there is broad consensus that many digital platforms exhibit significant tendencies toward ideological clustering \cite{cinelli2021echo}.

The causes of such clustering however remain contested, with two dominant competing explanations. One side of the debate emphasizes the role of automated curation, using the notion of `filter bubbles' \cite{pariser2011filter} to highlight the role of social media platforms' algorithmic designs in producing our online experience. In this view, recommender systems and news feeds -- especially on algorithmic-centric platforms like Facebook, TikTok and YouTube -- curate content in ways that reduce ideological diversity, often without users' awareness \cite{bakshy2015exposure,rader2015understanding}. While the aim is to offer users content that users are likely to appreciate, the unintended consequence is to insulate them from exposure to diverse viewpoints. 

The other side of the debate instead highlights the role of individual choice, suggesting that users engage in ``selective exposure'' \cite{stroud2010polarization} by avoiding dissonant topics and seeking to confirm their preexisting biases \cite{garrett2013turn}. The main culprit is not filtering algorithms, but users actively engaging in self-segregation, curating their own information environments by seeking out agreeable perspectives \cite{bakshy2015exposure}. Scholars argue that individuals gravitate toward ideologically aligned communities due to confirmation bias and social identity needs, independent of platform design \cite{sunstein2002law,barbera2015tweeting}. A growing body of empirical evidence supports this position: studies, some of which in collaboration with the platforms themselves, show that homophily is often stronger in user behavior than in algorithmic sorting \cite{colleoni2014echo}, that patterns of partisan engagement persist also when algorithmic amplification of political content is curtailed \cite{hosseinmardi2021examining}, and that deactivating algorithmic feeds has limited effects on polarization \cite{allcott2024effects}. Such findings have been taken to suggest that homogeneity is driven by intentional user choice rather than algorithmic filtering.

This paper challenges the foundation of this debate by presenting a simple agent-based model showing that high levels of ideological homophily can emerge in the absence of both filtering algorithms and user preferences. Using a simple model inspired by Schelling's \citep{schelling1971dynamic} model of residential segregation, we demonstrate that segregation can emerge as an unintended outcome driven by micro-level interactions triggering a self-reinforcing feedback loop that drives communities toward complete segregation. This dynamic suggests that echo chambers are not necessarily the product of neither deliberate user choices nor algorithmic design, but can instead be an emergent, system-level effect resulting from common interaction typologies. 

\section{A dynamic model of online segregation}\label{sec2}
We use as a starting-point Schelling's \citep{schelling1971dynamic} seminal model of residential segregation from 1971. Schelling's model is among the most influential in the social sciences, illustrating how individual-level preferences can produce large-scale social patterns without central coordination. In the model, agents of two types (e.g., red and blue, often interpreted as racial groups) occupy locations on a torus lattice, often taken to represent an urban area. At each time step, agents assess the composition of their local Moore neighborhood. If the proportion of same-type neighbors falls below a given tolerance threshold, they relocate to a randomly selected vacant site. This dynamic is repeated for a given number of steps, or until all agents all satisfied. Despite its simplicity, the model yields striking results: even when agents are tolerant of diversity --- requiring, for example, only 30\% of neighbors to be like themselves --- the system evolves toward near-complete segregation. This emergent macro-pattern, in which observed homophily far exceeds agents' preferences, is here referred to as the `Schelling segregation effect'.

The mechanism underlying this effect is a positive feedback loop: as agents move, they alter the composition of neighborhoods, triggering further dissatisfaction and migration. Although mixed configurations are theoretically stable, they are dynamically fragile and tend to collapse into polarized states. The model has since served as a canonical demonstration of how local decisions can unintentionally generate globally suboptimal outcomes and is so robust as to have been extended to a wide array of domains -- from social systems to phase separation in physics.

However, Schelling’s model is rooted in a spatial metaphor: agents interact with physical neighbors on a lattice. While this captures a broad range of systems, it poorly reflects the structure of online environments, where interactions typically occur in bounded groups or networks rather than spatially. The shift to new forms of social interaction is one of the most consequential transformations brought about by digital technology, with wide-ranging downstream effects—including a transition from normal to power-law distributions in many social phenomena. The question, then, is whether the core dynamic identified by Schelling generalizes beyond spatial systems to the typologies of digital platforms.

To explore this, we adapt the Schelling segregation model to capture the interactional dynamics of online communities while preserving its minimalistic structure. In our version, $N$ agents are distributed randomly across $C$ discrete communities, representing bounded digital environments such as online forums or chat groups. Each agent holds a fixed binary opinion, $o \in \{0, 1\}$, assigned at random.

At each time step, a focal agent is selected at random and interacts with $k$ other agents, sampled randomly from their current community. The agent then computes the share $f$ of interactors who hold the same opinion. If $f \leq \theta$, where $\theta$ is a threshold representing tolerance for disagreement, the agent becomes dissatisfied and relocates to a randomly selected new community. Crucially, agents do not change their opinion and have no intrinsic preference for any particular group -- only a weak homophily preference governs their decision to move.

Average final homophily, here denoted $\phi$, is quantified as the expected probability that a random within-community interaction occurs between agents holding opposing opinions. A segregation value of 0 indicates complete ideological sorting (i.e., no cross-opinion interactions within any group), whereas a value of 0.5 reflects maximal diversity (random mixing). If $\phi > \theta$, we can conclude that a Schelling segregation effect is in play.

This simple model allows us to examine whether homogeneity can emerge in digital settings absent both algorithmic curation and strong user preferences for like-mindedness -- purely as a result of the architecture and dynamics of online group interaction.

\begin{figure}[htbp]
  \centering
  \includegraphics[width=\linewidth]{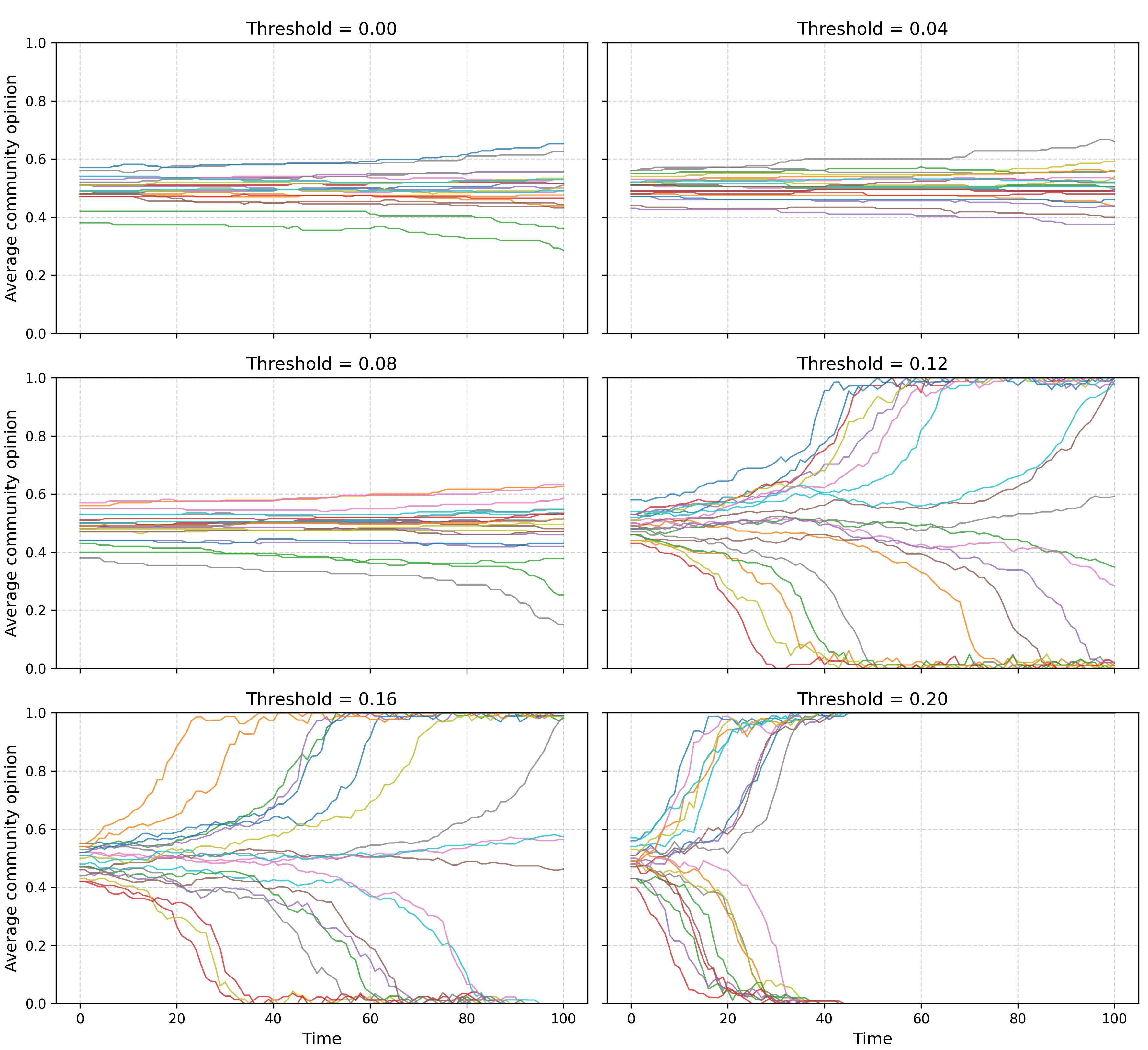}
  \caption{The evolution of the average opinions of each community as a function of $\theta$, i.e., the threshold by which the agents move if fewer than this number of their interactors are of the same type. Parameters are $N$=100, $C$=20, $k$=10. We run the model 100k steps; while the model does not necessarily reach a stable equilibrium during this time, this allows capturing the strength of the segregation mechanisms -- and whether it is likely to be realized. As can be seen, already when the agents require only 12\% of the interactors to be of their own type, nearly complete macro-level segregation results.}
  \label{fig:community_opinion_dynamics}
\end{figure}

\begin{figure}[htbp]
  \centering
  \includegraphics[width=0.7\linewidth]{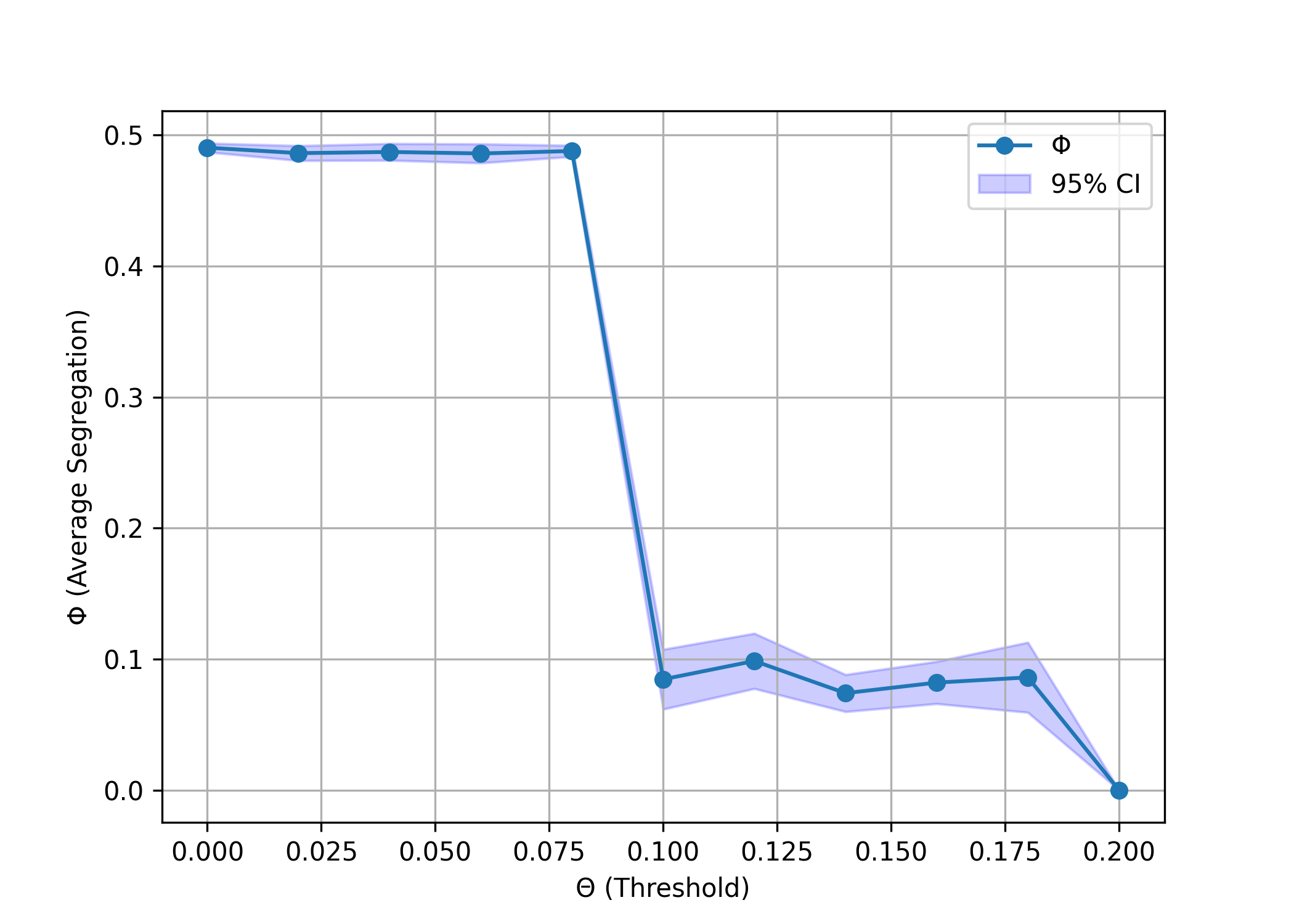}
  \caption{The final segregation value as a function of $\theta$, using the same parameters as Fig \ref{fig:community_opinion_dynamics}. The model was run 10 times for each data point, and the results are averaged over each run, here shown with 0.95 confidence intervals. The figure suggests that even when users require merely 10\% of their interactors to be of the same type as themselves, nearly complete segregation results. }
  \label{fig:segregation_by_threshold}
\end{figure}

\begin{figure}[htbp]
  \centering
  \includegraphics[width=0.5\linewidth]{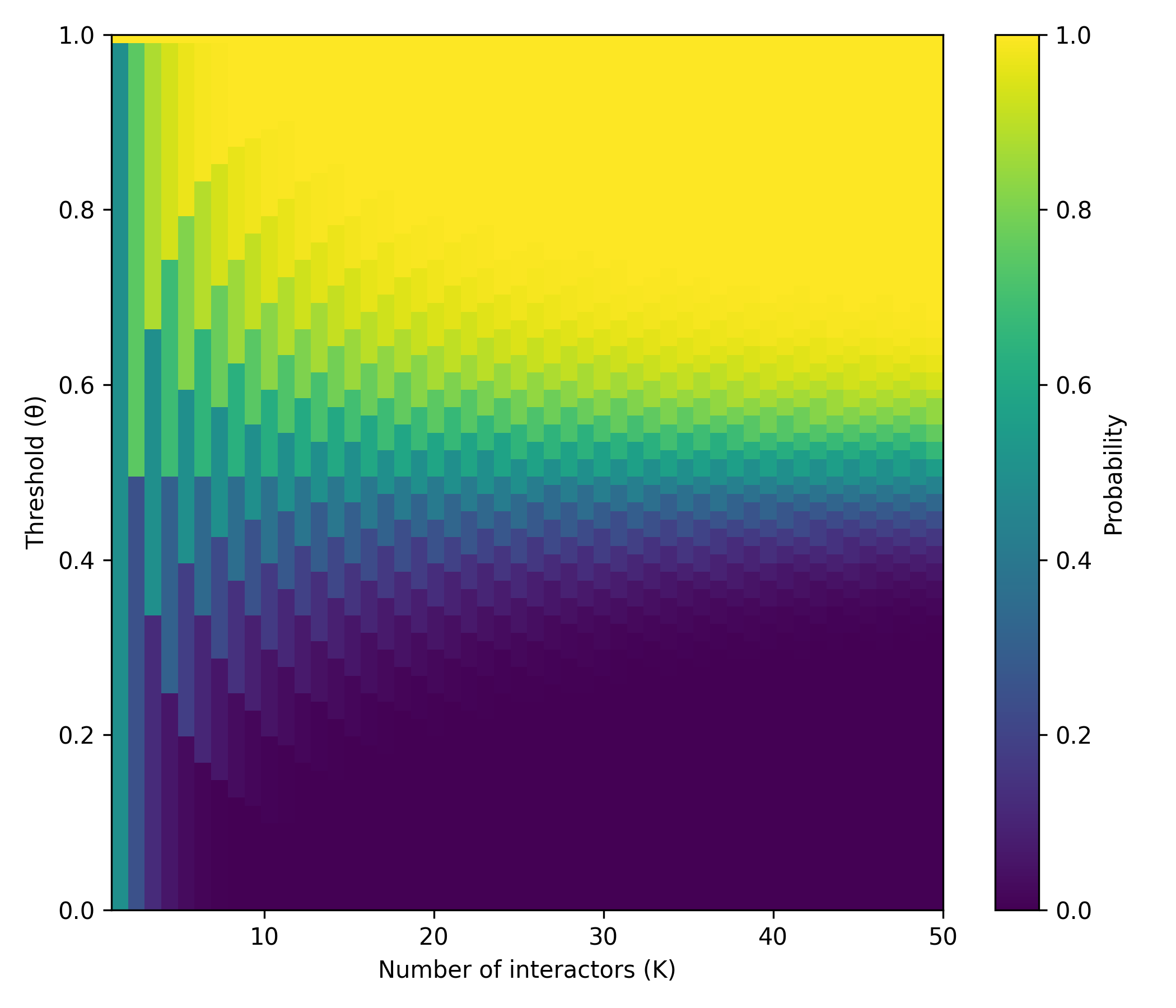}\includegraphics[width=0.5\linewidth]{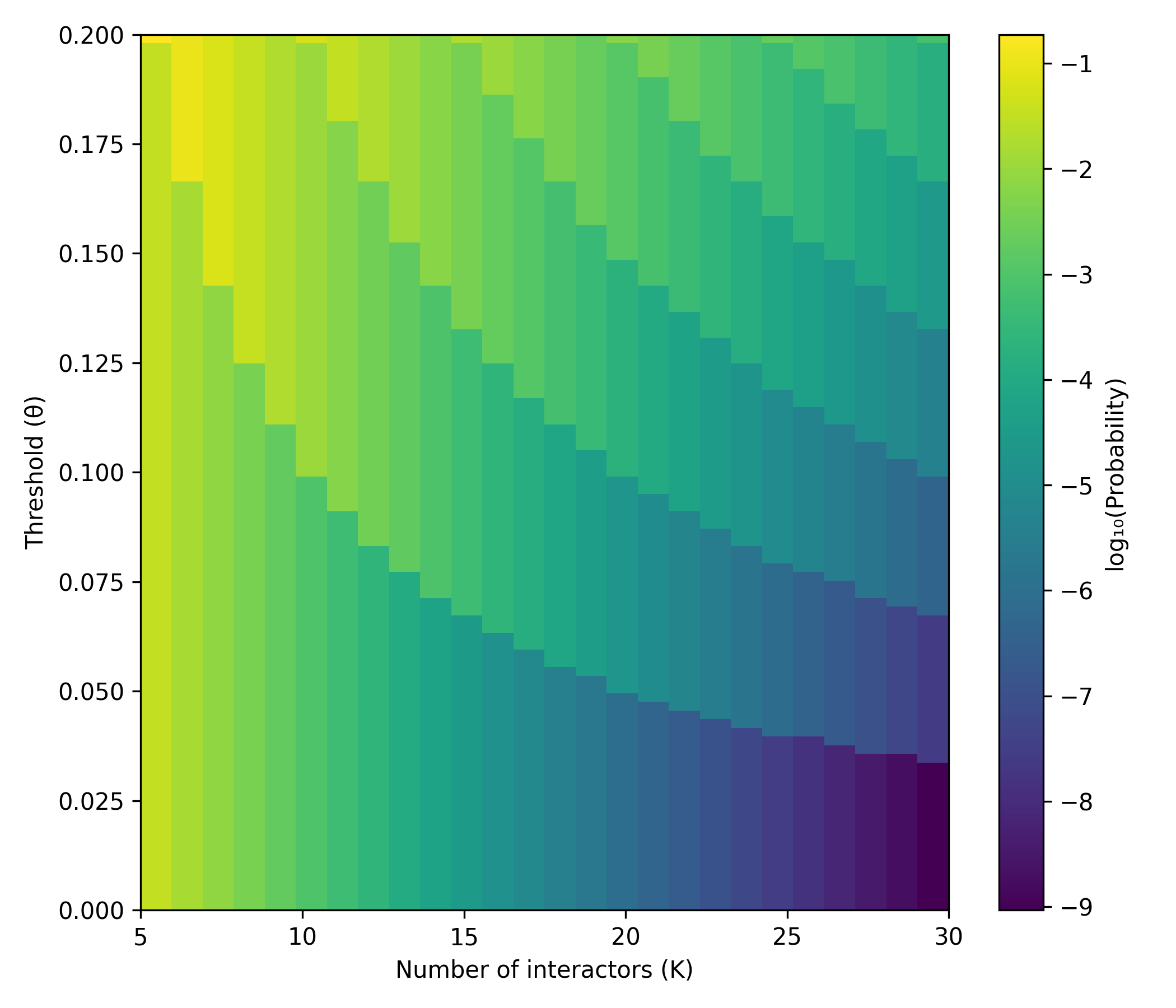}
  \caption{The analytical solution to the probability of an agent moving from a fully integrated community as a function $k$ and $\theta$. The right figure shows the log of the resulting probability, zooming in on the relevant part of the landscape. Given 100k step runs, a probability of around $10^{-2}$ is sufficient to enable a tipping point and drive complete segregation.}
  \label{fig:binomial_probabilities}
\end{figure}

\section{Results}
Fig~\ref{fig:community_opinion_dynamics} presents the temporal evolution of average community opinion across varying values of the threshold parameter \( \theta \). When \( \theta=0 \) or is very low, agents are almost never dissatisfied with their group composition, and no meaningful segregation emerges. However, as \( \theta \) increases, a phase transition becomes evident. At \( \theta = 0.12 \), the system begins to exhibit strong polarization, and at \( \theta = 0.2 \), complete segregation occurs rapidly, often within just a few steps. This pattern illustrates how even modest increases in intolerance to dissimilar opinions can produce starkly different system-level outcomes.

Fig~\ref{fig:segregation_by_threshold} shows the final segregation level \( \phi \) as a function of \( \theta \). The results reveal a sharp transition around \( \theta = 0.1 \), beyond which the system consistently converges to highly segregated states. Notably, this tipping point is substantially lower than in the classical spatial Schelling model, where segregation typically emerges at tolerance levels around 0.3 \cite{schelling1971dynamic}. In our networked community model, segregation emerges at significantly weaker homophily thresholds, indicating that the online community-based interaction structure amplifies the Schelling segregation effect.

To understand this dynamic analytically, consider the probability that a user will migrate when placed in a fully integrated community -- where opinions are equally distributed. This situation can be modeled as a binomial probability: the agent samples \( k \) peers and will move if the fraction of similar opinions $H$ is less than or equal to \( \theta \). Formally, the probability of migration is given by:

\[
\Pr\left(\text{migrate} \right) = \Pr\left(H \leq \left\lfloor \theta \cdot k \right\rfloor \right) = \sum_{i=0}^{\left\lfloor \theta \cdot k \right\rfloor} \binom{k}{i} (0.5)^i (0.5)^{k-i} = \sum_{i=0}^{\left\lfloor \theta \cdot k \right\rfloor} \binom{k}{i} (0.5)^k
\]

Fig~\ref{fig:binomial_probabilities} plots this probability as a function of \( k \) and \( \theta \). The graph reveals a discrete jump in migration probability around \( \theta = 0.1 \) for \( k = 10 \). Specifically, when \( \theta < 0.1 \), the agent will only move if they observe no similar opinions—yielding a migration probability of \( 0.00098 \). When \( \theta > 0.1 \), observing just one similar peer is insufficient, raising the migration probability to \( 0.01074 \)—an order of magnitude increase. This nonlinear jump activates a critical cascade: agents who find themselves in a slight minority are more likely to relocate, incrementally shifting the composition of both origin and destination communities. These small imbalances compound through a positive feedback loop—each departure increases dissatisfaction for remaining minority agents, while randomly reinforcing homogeneity in other groups. Over time, this process results in nearly complete ideological sorting, even when initial conditions are fully integrated and individuals prefer integrated states.

\section{Discussion}
This study has demonstrated that the Schelling segregation effect not only generalizes beyond spatial environments to group-based interaction structures but in fact becomes more pronounced in these settings. Our model reveals how even mild preferences for homophily can trigger tipping dynamics that drive communities from integrated to segregated states. Once a slight majority of like-minded individuals forms in a group, it increases the likelihood that minority users will become dissatisfied and relocate. Their departure reinforces homogeneity in the original group and may seed new homogeneous clusters elsewhere, initiating a self-reinforcing feedback loop. The result is a bifurcation of the system into internally uniform and mutually alienated communities.

Crucially, these dynamics imply that small initial imbalances can determine a group's future trajectory. For example, a benign interest-based community -- such as a male-dominated gaming forum -- might, through repeated cycles of departure and resettlement, evolve into an ideologically extreme enclave \cite{phelan2024radicalisation,bezio2018ctrl}. This mechanism offers a plausible theoretical underpinning for observed trajectories of gradual radicalization within closed online spaces.

The model has two important and counterintuitive implications for research on online homophily and `echo chambers'. First, homogeneous online communities can emerge even in the absence of algorithmic filtering or explicit user preferences for ideological alignment. Online homogeneity may be neither designed nor desired; rather, it can emerge as unintended outcomes of seemingly benign interaction structures. This perspective reframes online homogeneity as an emergent, system-level phenomenon, rather than the product of deliberate individual or platform behavior.

Second, algorithmic filtering -- so-called `filter bubbles' -- may paradoxically \textit{reduce} homophily. By increasing perceived similarity within communities -- e.g., by boosting visibility of like-minded users -- platforms might suppress the dissatisfaction that drives relocation. While algorithmic curation is often blamed for insulating users, it may in some cases help preserve heterogeneity within groups by preventing them from tipping into polarized states.

A natural question concerns whether the Schelling segregation effect extends also to network-based interaction structures, such as those found on Twitter/X, Bluesky, or Facebook. We can conclude that this is unlikely to be the case. The segregation cascade observed in our model relies on transitivity: when one user leaves a group, it changes the local environment for both the group that they leave and the group that they join, increasing the likelihood of others leaving. In networks, unfollowing a user alters one's personal feed -- and may in cases of undirected ties affect the other user's feed -- but it does not make anyone more likely to become dissatisfied with their network. For Schelling-like dynamics to occur in a network, agents would need to abandon entire subnetworks and migrate to new areas in the network, which would moreover have to be highly transitive. The network would, in short, have to embody a form of group-based or spatial interaction structure -- for which there is little empirical or theoretical basis. Therefore, while networked environments can still exhibit homophily, they are unlikely to produce more homogeneity than users desire. This aligns with empirical studies showing greater interactional diversity on networked platforms like Twitter/X than on group-based platforms such as Reddit or Telegram \cite{cinelli2021echo, bakshy2015exposure}.

\section{Conclusion}
This paper contributes to the study of online homogeneous spaces by applying one of the foundational insights from complexity science: micro-level preferences and behaviors, when aggregated, can produce emergent macro-level patterns that were neither designed nor intended by any single individual. Through a minimal agent-based model inspired by Schelling’s classic work, we demonstrate how even weak homophily, absent any centralized design or strong ideological motivations, can give rise to sharply segregated online communities.

The results highlight that polarization in digital spaces may not require neither extreme preferences, malign intent, nor algorithmic manipulation. Instead, it can result from simple interaction rules playing out over time across bounded community structures. This insight provides a theoretical foundation for understanding how ideologically uniform enclaves form -- even when most users prefer diversity. 

At the same time, it is important to highlight the limits of abstraction. Models are necessarily simplifications; they identify potential social mechanisms but do not guarantee their realization. In practice, digital users possess multifaceted identities, engage across diverse groups, and interact on issues of varying salience. These complexities may moderate or delay the emergence of segregation. 

The model presented in this brief paper has offered a metaphor for platform governance: maintaining ideological diversity within online communities may resemble the delicate act of balancing a stick on one's palm. It is achievable -- but demands ongoing, constant correction to prevent the system from tipping over into polarization. At the same time, lessons from complexity underscored how interventions in complex socio-technical systems can yield unintended -- and at times counterproductive -- consequences, as emergent dynamics may subvert even well-intentioned designs. 

\section*{Code Availability}
The complete code for the model is available at \url{http://github.com/cssmodels/onlinesegregation}

\bibliography{sn-bibliography}% common bib file

\end{document}